\def\preprint{1}                
\def\comment#1{}

\if\preprint1
        \documentclass[usenatbib]{mn2e}
        \usepackage{natbib}
        \usepackage{times}
        \usepackage{graphicx}
        \usepackage{calc}
        \usepackage{lscape}

\else
        \documentstyle[astrop-bib,referee,times]{mn}
        \newcommand{\includegraphics}[1]{}
\fi

\def\oversim#1#2{\lower0.5pt\vbox{\baselineskip0pt \lineskip-0.5pt
     \ialign{$\mathsurround0pt #1\hfil##\hfil$\crcr#2\crcr\sim\crcr}}}
\def\gsim{\mathrel{\mathpalette\oversim>}}    
\def\lsim{\mathrel{\mathpalette\oversim<}}    

\title[Magnetic fields in PNe]
{Magnetic fields in planetary nebulae and post-AGB nebulae}
\author[L. Sabin et al.]
       {L. Sabin$^{1,2}$, Albert~A.~Zijlstra$^1$ and J.~S.~Greaves$^3$\\
        $^1$Jodrell Bank Centre for Astrophysics, University of 
        Manchester, P.O. Box 88, Manchester M60 1QD, UK\\
        $^2$Instituto de Astrof$\acute{\imath}$sica de Canarias, C/ V$\acute{\imath}$a L$\acute{a}$ctea,
        38205 Laguna, Tenerife, Canary Islands, Spain\\
       $^3$School of Physics \&\ Astronomy, University of St. Andrews,
        North Haugh, St Andrews KY16 9SS, Scotland, UK}

\pubyear{2006}

\begin{document}

\maketitle

\begin{abstract}

Magnetic fields are an important but largely unknown ingredient of planetary
nebulae. They have been detected in oxygen-rich AGB and post-AGB stars, and
may play a role in the shaping of their nebulae. Here we present SCUBA
sub-millimeter polarimetric observations of four bipolar planetary nebulae and
post-AGB stars, including two oxygen-rich and two carbon-rich nebulae, to
determine the geometry of the magnetic field by dust alignment.  Three of the
four sources (NGC 7027, NGC 6537 and NGC 6302) present a well-defined toroidal
magnetic field oriented along their equatorial torus or disk.  NGC 6302 may
also show field lines along the bipolar outflow.  CRL 2688 shows a complex
field structure, where part of the field aligns with the torus, whilst an
other part approximately aligns with the polar outflow.  It also presents
marked asymmetries in its magnetic structure.  NGC 7027 shows evidence for a
disorganized field in the south-west corner, where the SCUBA shows an
indication for an outflow.  The findings show a clear correlation between
field orientation and nebular structure.

\end{abstract}


\section{Introduction}

Whether magnetic fields play a role in the shaping of planetary nebulae (PNe)
is an open question.  Most of the post-AGB nebulae appear elliptical, bipolar
or even multi-polar \citep{Balick2002}.  These morphologies are amplified by
the interaction between a slow AGB wind with a faster post-AGB wind. However,
this amplification still requires an initial asymmetry in the slow wind. This
initial shaping has been mainly attributed to two possible phenomena: binarity
and magnetic fields.\\

In the binary model, a close companion affects the mass-losing AGB star via
common envelope evolution \citep{DeMarco2006}, mass transfer and/or tidal
forces, and may lead to the formation of an accretion disk around the
companion star. The binary orbit provides a source of angular momentum which
get carried by the wind, and leads to an equatorial disk. The angular momentum
loss by the stars may lead to a merger. The model of shaping due to this
binary interaction is quite popular \citep{Bond2000}, \citep{Ciardullo2005},
partly because of the presumed impossibility for a single star to supply the
energy necessary to create a magnetic field strong enough for its shaping
\citep{Soker2005}.  Nevertheless, there is still a lack of observational
evidence for the occurrence of close binary systems during the AGB
phase. Neither companions \citep{Riera2003} nor high orbital velocities of the
AGB stars \citep{Barnbaum1995} are detected in a sufficient amount to
establish the role of the binarity as predominant \citep{Matt2000}. Evidence
for binary interactions may be found in 25--50 per cent of planetary nebulae
\citep{Zijlstra2006}, although  values of 50--100 per cent have also
been suggested  \citep{DeMarco2004, DeMarco2006}.\\

On the other hand, the magnetic field may act as a ``squeezer'' around the
central star of the PN and thereby give  the dust its direction (in the
sense of the outflow). Magnetic fields have been detected around AGB stars
\citep{Vlemmings2006} and a few post-AGB stars \citep{Bains2004} using radio
observations of masers (H$_2$O, SiO, OH). Such observations measure a local
value of the strength of field, within masing high-density clumps, which may
differ from the global field.  The origin of the magnetic field is unknown: a
dynamo effect resulting from an interaction between a slow rotating envelope
and a fast rotating core has been proposed \citep{Blackman2001}.  As magnetic
fields are now known to be present in the AGB and the post-AGB phase, 
their importance should not be ruled out prematurely.

\cite{Greaves2002} found evidence for dust alignment in two carbon-rich
objects, NGC 7027 and CRL2688 which are respectively a young and compact
bipolar PN and a strongly bipolar proto-planetary nebula (PPN).  This was
based on polarimetric Scuba 850-micron observations. The data suggests the
presence of toroidal collimated magnetic fields, as would be required for the
shaping.  But the presence of such a field was not conclusively proved,
because of the very few detected vectors and limited spatial resolution.  We
present here new 850-microns and the first 450-microns polarimetry of post-AGB
stars, which allows for better resolution and reduces angular smearing.  In
addition to the objects observed by \cite{Greaves2002}, we also observed NGC
6537 and NGC 6302. The sample contains two carbon-rich and two oxygen-rich
nebulae. We show the presence of well-aligned toroidal fields in three of the
nebulae. The fourth object shows indications for both a toroidal field and one
aligned with the polar outflow. We conclude that toroidal fields may be common
in bipolar PNe, and could play a role in the shaping of the nebulae.

\section{Observations}

The polarimetric data have been obtained  May 10$^{\rm th}$ 2005, with the
polarimeter on the Sub-millimeter Common-User Bolometer Array (SCUBA), at the
JCMT. The instrument (now decommissioned) is described in \citet{Holland1999}.
The JCMT beam size is 15 arcsec at 850$\mu$m and 8 arcsec at 450 $\mu$m.

SCUBA contains two arrays of bolometric detectors, covering a field of view of
2.3$^\prime$ in diameter. The 850-$\mu$m array has 37 individual detectors, and
the 450-$\mu$m array has 91 detectors.  The two arrays are used
simultaneously. The spatial resolution of the detector is 7.5'' at
450\,$\mu$m, and 14'' at 850\,$\mu$m. The gaps between the detectors are
covered by moving the telescope in sub-pixel steps, in the so-called
jiggle-map mode.  The step size needs to be optimized for the wavelength used.
The polarimeter \citep{Greaves2003} measures the linear polarization by
rotating the half-waveplate in 16 steps. We used this in combination with
jiggle map mode. A chop-throw of 45 arcsec was used.

Good photometric images cannot be obtained at both wavelengths
simultaneously. When the jiggle pattern is optimized one wave-band, the
simultaneous image obtained in the other band is under-sampled.  The size of
the pixels in the final image is set during the data reduction with the
polarimetric package of Starlink.  For a jiggle pattern corresponding to a
450$\mu$m measurement, the pixel spacing becomes 3 arcsec and for 850$\mu$m it
is 6 arcsec. The reduction of infrared polarimetry is discussed in detail by
\cite{Hildebrand2000}.

The instrumental polarization (IP) of each detector has to be removed in order
to have a correct polarization calibration.  A new IP calibration, provided on
site, was used: this contains a re-measurement of the central detector.  For
the other detectors, an older IP calibration was used.  The IP should not vary
too much for detectors close to the center of the array. Some tests done to
see the importance of the instrumental polarization (by changing its value by
the size of the error for example ) and if it could affect the polarization
vectors length, showed that in our case, the IP was small (at 850$\mu$m:
$\sim$1.20\%$\pm$0.25\% and at 450$\mu$m: $\sim$3.25\%$\pm$0.25\% ) and didn't
play any role in modifying the vectors.  The resulting polarimetric images
were checked for different pixel binning, from 1$\times$1 to 10$\times$10, and
we didn't find any significant changes. 

 The linear polarization is measured as a percentage polarization, and a
direction. The polarization is typically caused by the alignment of spinning
dust grains, with their long axis perpendicular to the local magnetic field
\citep{Greaves1999}. Thus, the measured angle of polarization is  90 degrees 
rotated with respect to the magnetic field.  The degree of polarization
does not give direct information on the strength of the magnetic
field. 

We observed four targets: NGC 6537, CRL 2688, NGC 6302 and NGC 7027. They were
observed in jiggle map observing mode at both wave-bands, 450$\mu$m and
850$\mu$m.  For each object, the mean direction and angle of polarization are
listed in Table 1. 

The figures below show the direction of the magnetic field, i.e. the vectors
are perpendicular to the direction of the grain alignment.

\section{The results}

\subsection{NGC 6537}

NGC 6537 is also known as the Red Spider nebula. It is a bipolar planetary
nebula with a very hot central star (1.5--2.5$\times$$10^5$K). The nebula
suffers extinction by circumstellar dust. This extinction is localized mainly
in a compact circumstellar shell. An extinction map \citep{Matsuura2005}
reveals a compact dust shell with a roughly spherical inner radius of about 3
arcsec, with a minimum towards the central star \citep{Matsuura2005,
Cuesta1995}. The polar outflows \citep[traced by high velocity winds of about
300km/s:][]{Corradi1993} extend 2 arcminutes along the NE-SW direction. The
Scuba 850-$\mu$m continuum map (Fig. 1) shows elongation perpendicular to the
outflow direction, with a major-axis diameter of approximately 20 arcsec. The
extinction map is now seen to represent the inner edge of a more extended, and
possibly toroidal, structure. The equatorial plane may be oriented a little
closer to the EW direction, based on the Scuba map. \\

The best polarimetric results obtained with SCUBA are obtained at
850$\mu$m. The consistent orientation of the polarization vectors shows that
the magnetic field (hereafter $\vec{B}$) has a dominant direction along the
equatorial plane, approximately perpendicular to the outflow direction. The
dust alignment is therefore directed in the same sense as the outflow. The
length of the 18 different vectors does not show large variations, with a
degree of polarization varying from 8 to 14$\%$, suggesting a consistent
magnetic field. Moreover the absence of smaller polarization vectors toward
the center of the nebula indicates that there is no change in geometry of
$\vec{B}$ towards the core \citep{Greaves2002}.  This supports a location some
distance from the star (i.e. in a detached shell), since otherwise averaging
of vectors in different directions within the JCMT beam would reduce the
detected net polarization at the central position (beam depolarization).

These observations indicate the presence of a consistent toroidal magnetic
field, located along the equatorial plane of NGC 6537, in a
circumstellar torus. Compared to the size of the outflow lobes, the field is
located relatively close to the star.  The presence of a $\vec{B}$-field had
already been suspected, based on the occurrence of filaments near the central
star \citep{Huggins2005}.

The extinction map obtained by \citet{Matsuura2005} (Fig.~1-Bottom panel)
shows an asymmetry, in that the extinction is higher on the western side
($\gsim$2 mag versus $\lsim$1.6 mag towards the east). This asymmetry is also
seen in the Scuba data, with a larger extension and more polarization vectors
on this side. This is a further indication that the magnetic field is located
within the detached dust shell.

There is no strong indication for magnetic fields along the spider lobes.
There is a slight trend for the vectors to curve, but this is caused by 
only a few of the vectors and would need confirmation. We cannot
state with confidence whether the lobes also carry a magnetic field.

\begin{figure}
\begin{center}
{\includegraphics [height=6.5cm]{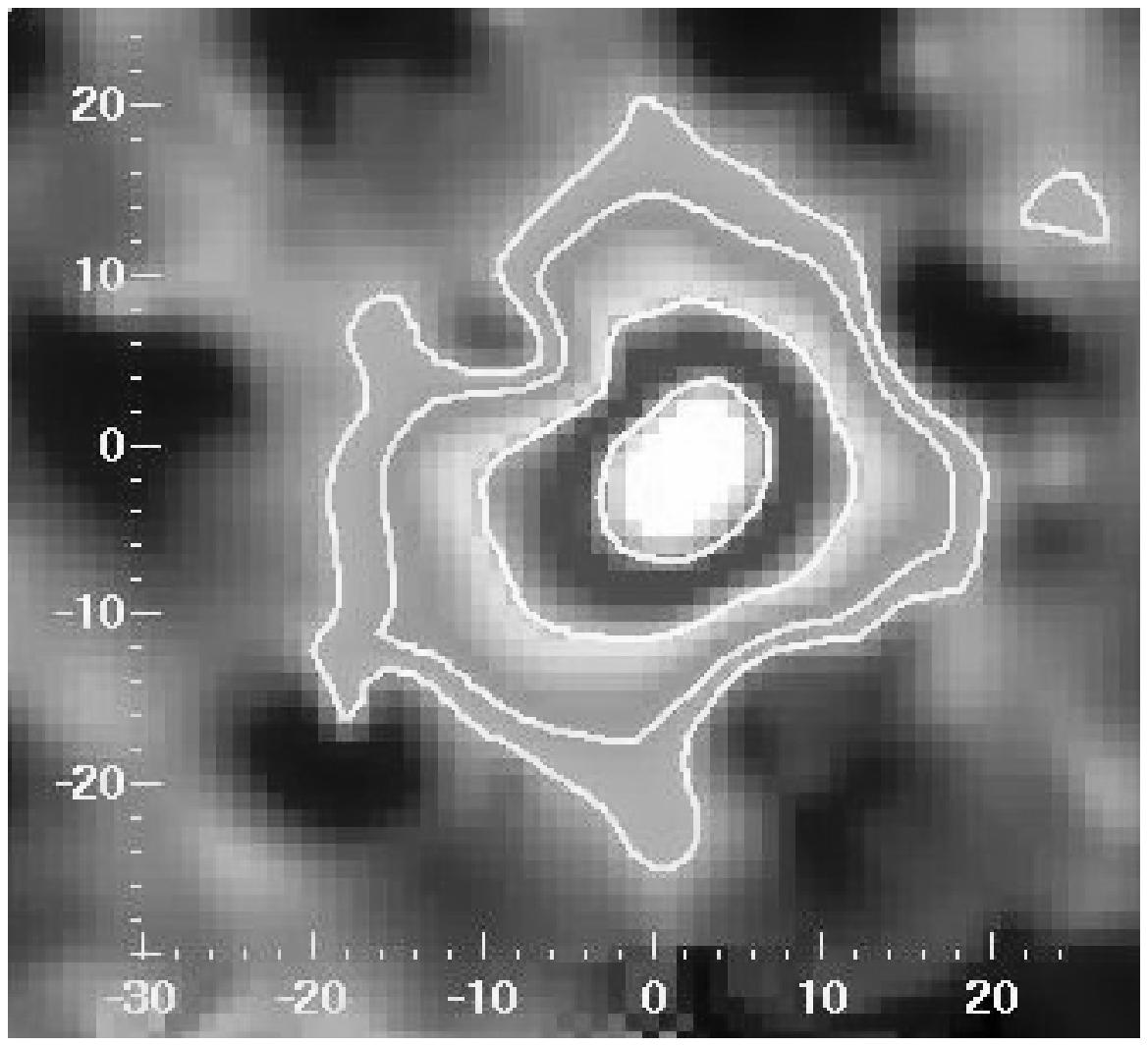}}
{\includegraphics [height=6.7cm]{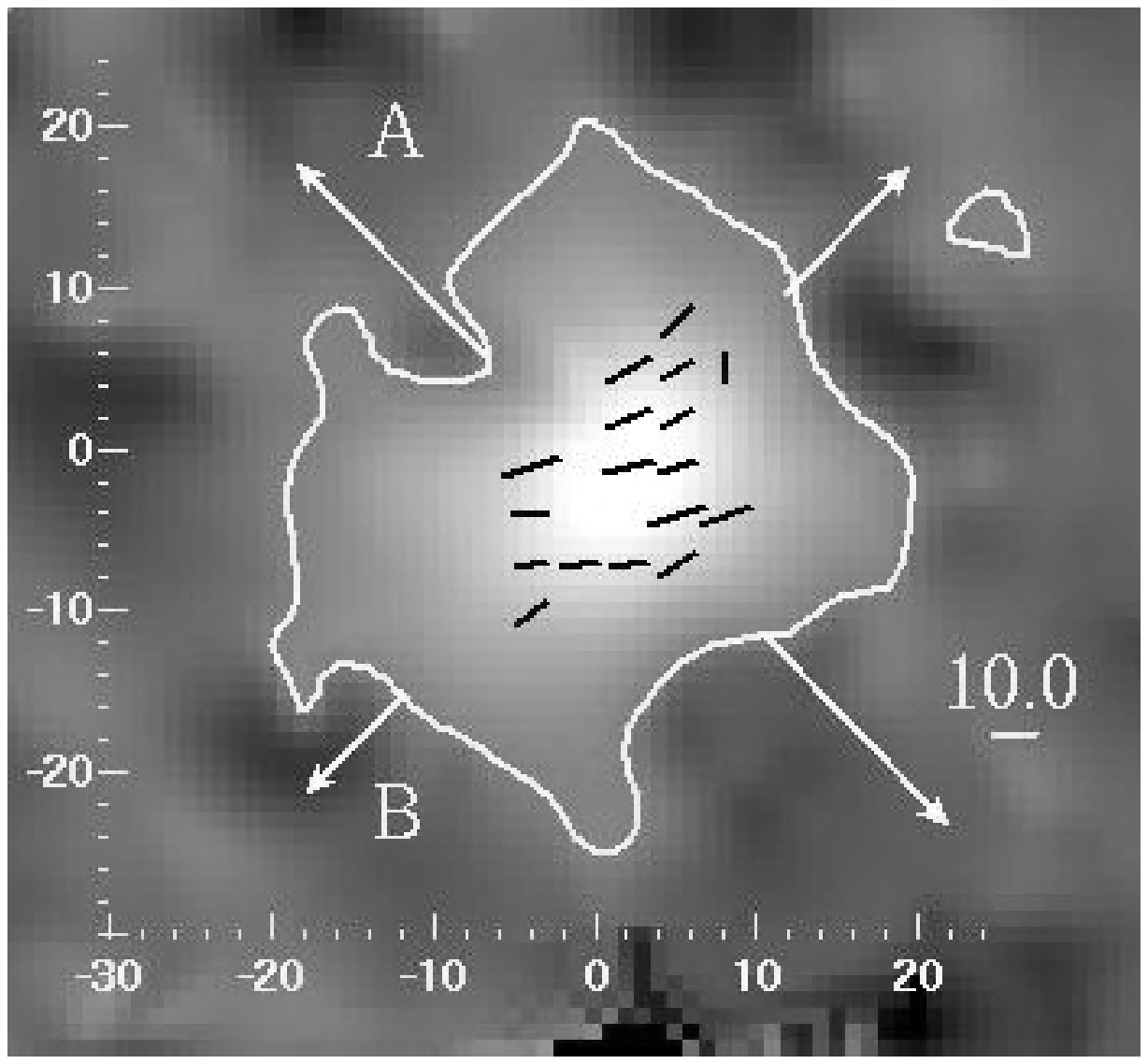}}
{\includegraphics [height=6.6cm]{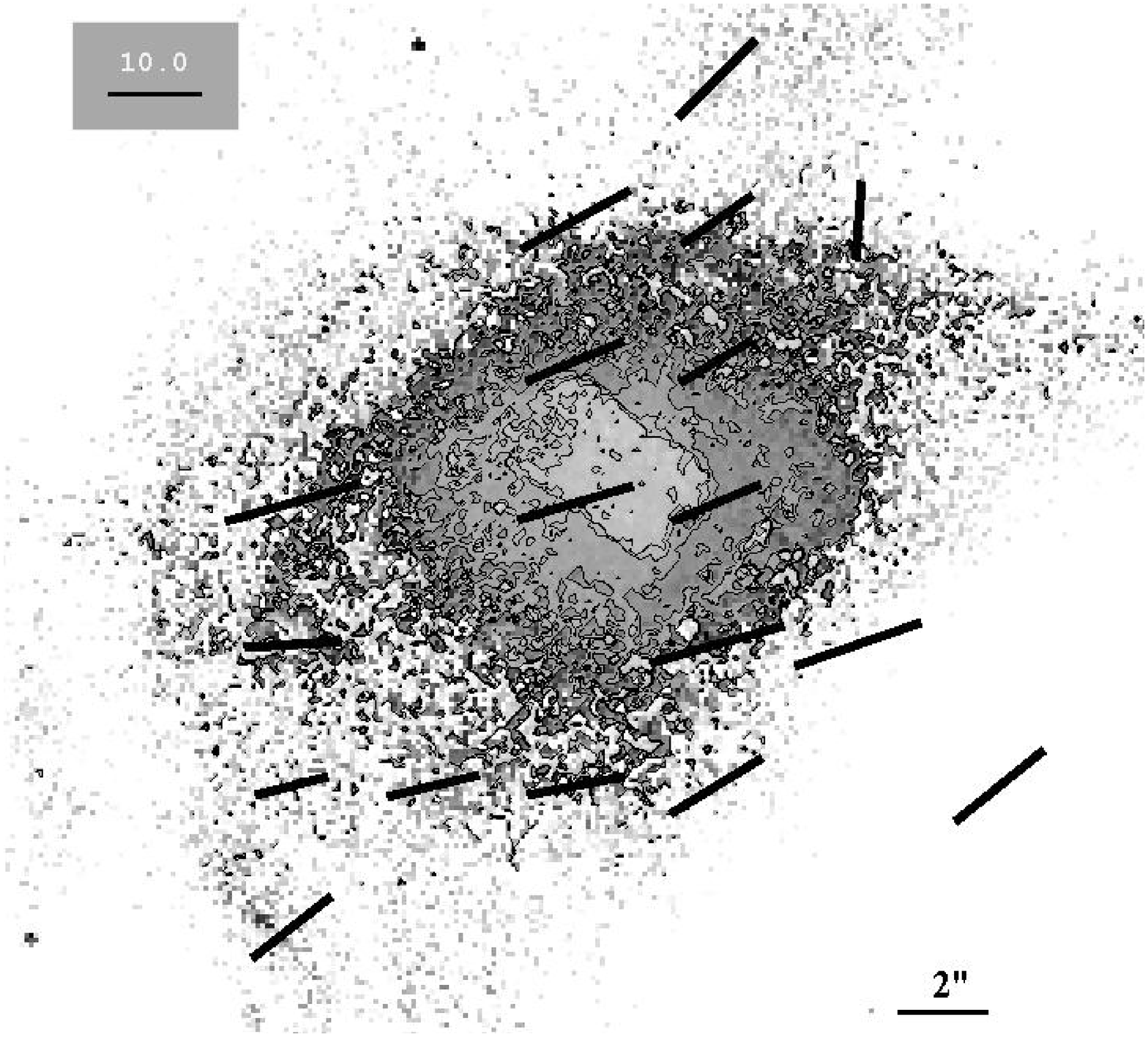}}
\caption{\label{}Scuba 850$\mu$m results on NGC 6537.  North on the top and
East on the left. The axes give the image scale in arcseconds.  {\it Top
panel}: the 850$\mu$m continuum map of NGC 6537, with contours at 1\%, 2\%
,5\% and 10\% of the peak. {\it Middle panel}: Magnetic field orientation. The
general outflow direction of the nebula is indicated by A and the equatorial
plane by B. The polarization vector scale on the left is set at 10\%. {\it Bottom panel}: Extinction map
presented by Matsuura et al. (2005). The highest levels of extinction occur at
$\sim$ 4 arcsec from the central star with $E\rm (H\beta-H\alpha) > 2\,mag$,
coincident with the dust emission in the 850$\mu$m map. The polarization
vector scale is set at 10\%. The magnetic field which coincides
with the area of internal extinction, is mostly aligned along the equatorial
plane, indicating a toroidal field. The dust alignment is perpendicular
to the vectors displayed. 
}
\end{center}
\end{figure}

\subsection{NGC 7027}

\begin{figure}
\begin{center}
{\includegraphics [height=6.9cm]{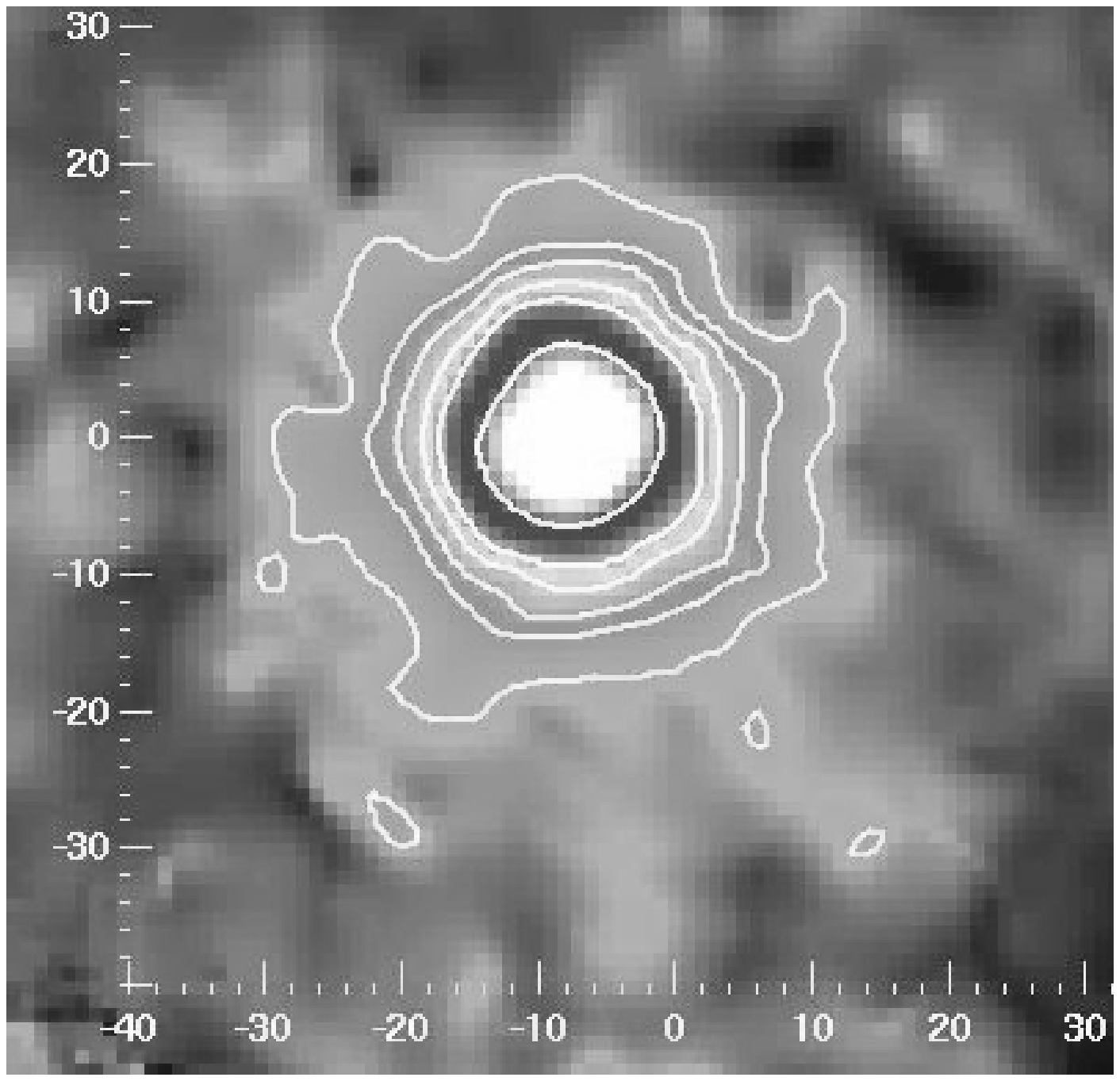}}
{\includegraphics [height=7.1cm]{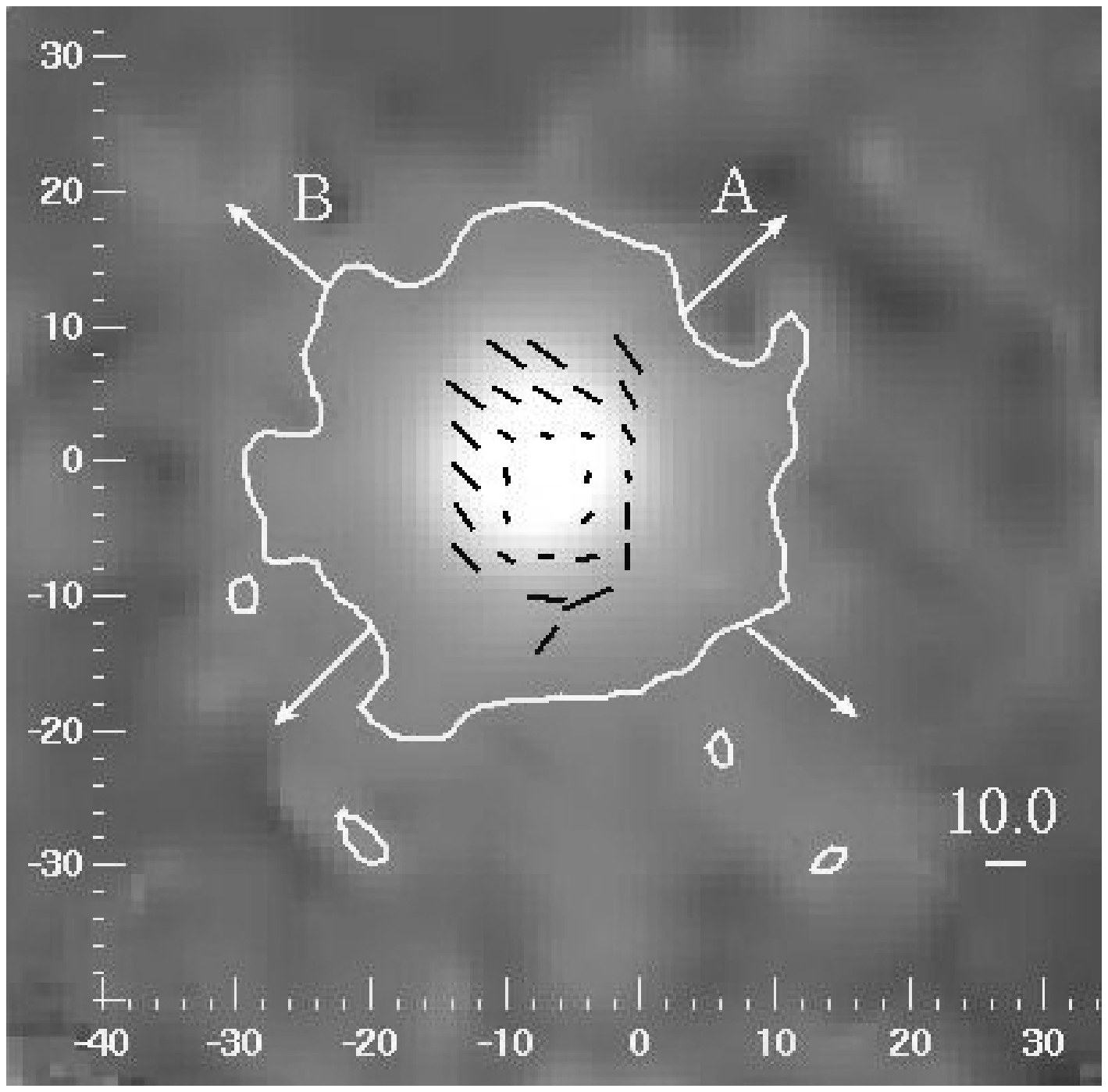}}
{\includegraphics [height=5.9cm]{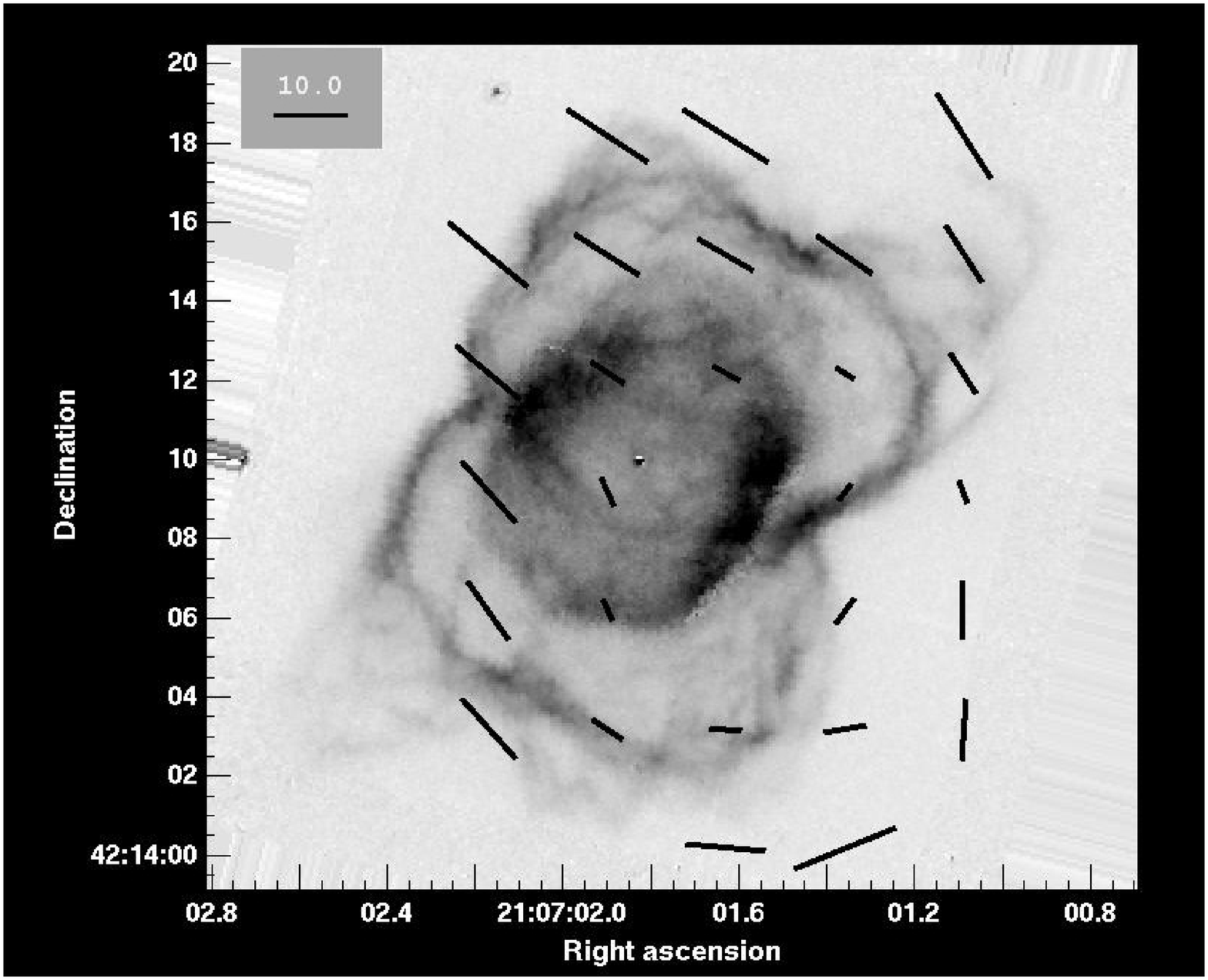}}
\caption{\label{}Scuba 450$\mu$m results on NGC 7027.  North on the top and
East on the left. The axes give the image scale in arcseconds.  {\it Top
panel}: The 450$\mu$m continuum map of NGC 7027. Contours are set from 1 to
5\% and 10\% of the peak. {\it Middle panel}: Magnetic field distribution.
The general outflow direction is indicated by A and the equatorial plane by
B. The polarization vector scale (showing the degree of polarization) is set
at 10\%.  {\it Bottom panel}: The Scuba polarization vectors are shown on an
continuum subtracted H$_2$ (color-inverted) map of NGC 7027 (North on the top
and East on the left). The field is mostly directed along the equatorial
plane. The central region shows much reduced polarization and the field
orientation differs on the extreme western side. 
}
\end{center}
\end{figure}

The 450$\mu$m jiggle map of the young planetary nebula NGC 7027 (Fig.~2) and
its near environment covers a field of about 40$\times$36 arcsec$^2$. The
central star is surrounded by an ionized area of $\sim$283 arcsec$^2$, which
is in turn enclosed by a thin atomic and molecular layer that is seen for
instance in the H$_{2}$ emission observed by \cite{Cox2002}.  A thin dark ring
delineates the equatorial plane in optical images.  The structure is better
seen in the HST image (Fig.~2-bottom), which shows both the surrounding
molecular layer (in H$_2$) and the inner ionized region. \cite{Latter2000}
shows that the NW lobe is blue-shifted (closer) and the SW lobe is
red-shifted.

The dominant direction of the magnetic field coincides with the equatorial
plane. But this behaviour is mainly seen on the North-East side while the
south-west part seems to be disturbed: the magnetic field may be ``broken''.
The degree of polarization is $8.9\% \pm 0.9\%$) towards the NE direction (or
lobe) and $ 7.6\%\pm 1.3\%$) towards the SW. The degree of polarisation is
much reduced in the center of the nebula and lacks a coherent direction here.
This effect was also noted by \citet{Greaves2002} and may indicate that
coherence is lost in the ionized region.

\subsection{CRL 2688}
\begin{figure*}
\begin{center}
\hspace{5cm}
\hbox{
{\includegraphics [height=9cm]{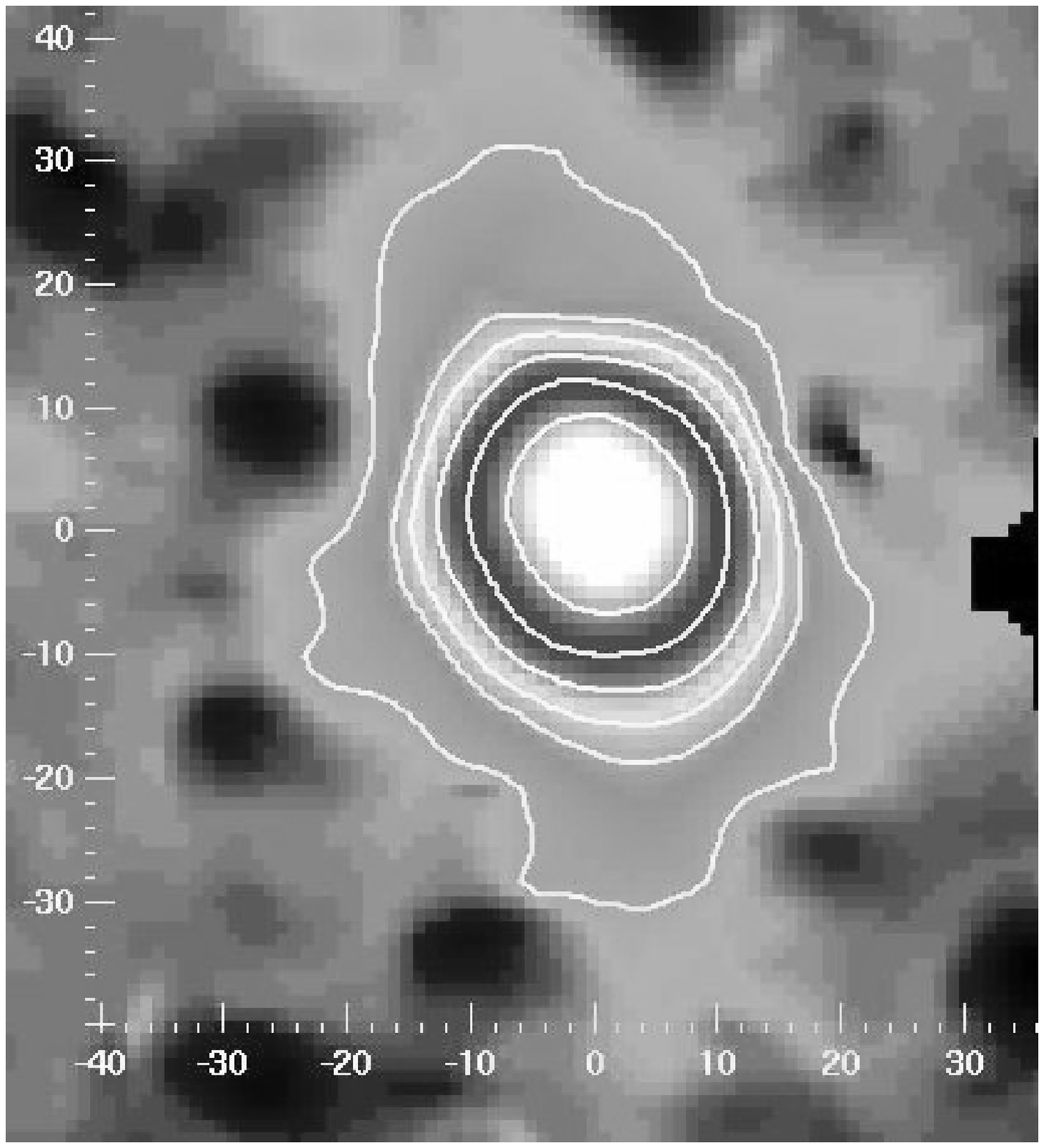}}
{\includegraphics [height=9cm]{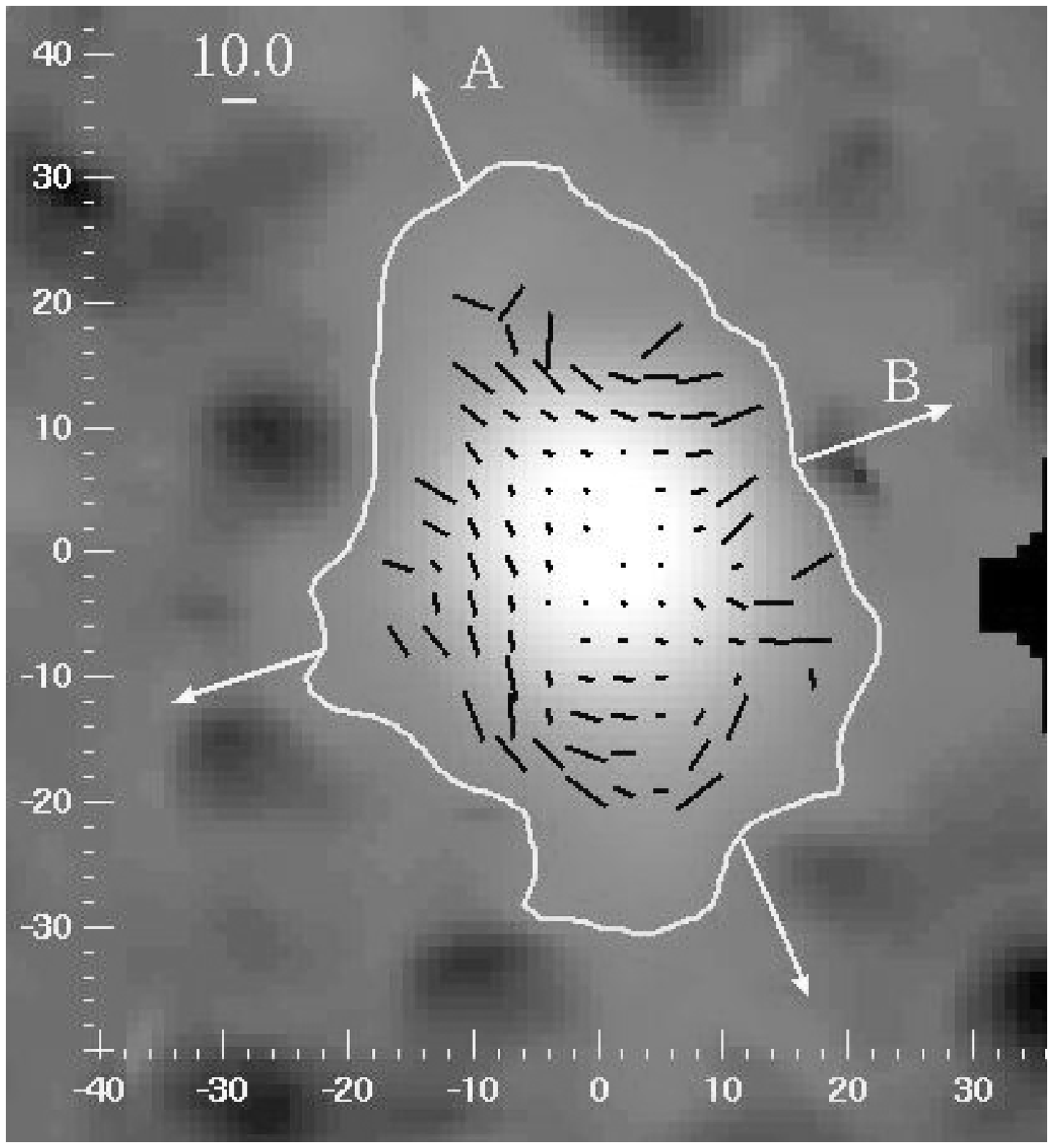}}
}
\hbox{
{\includegraphics [height=7.5cm]{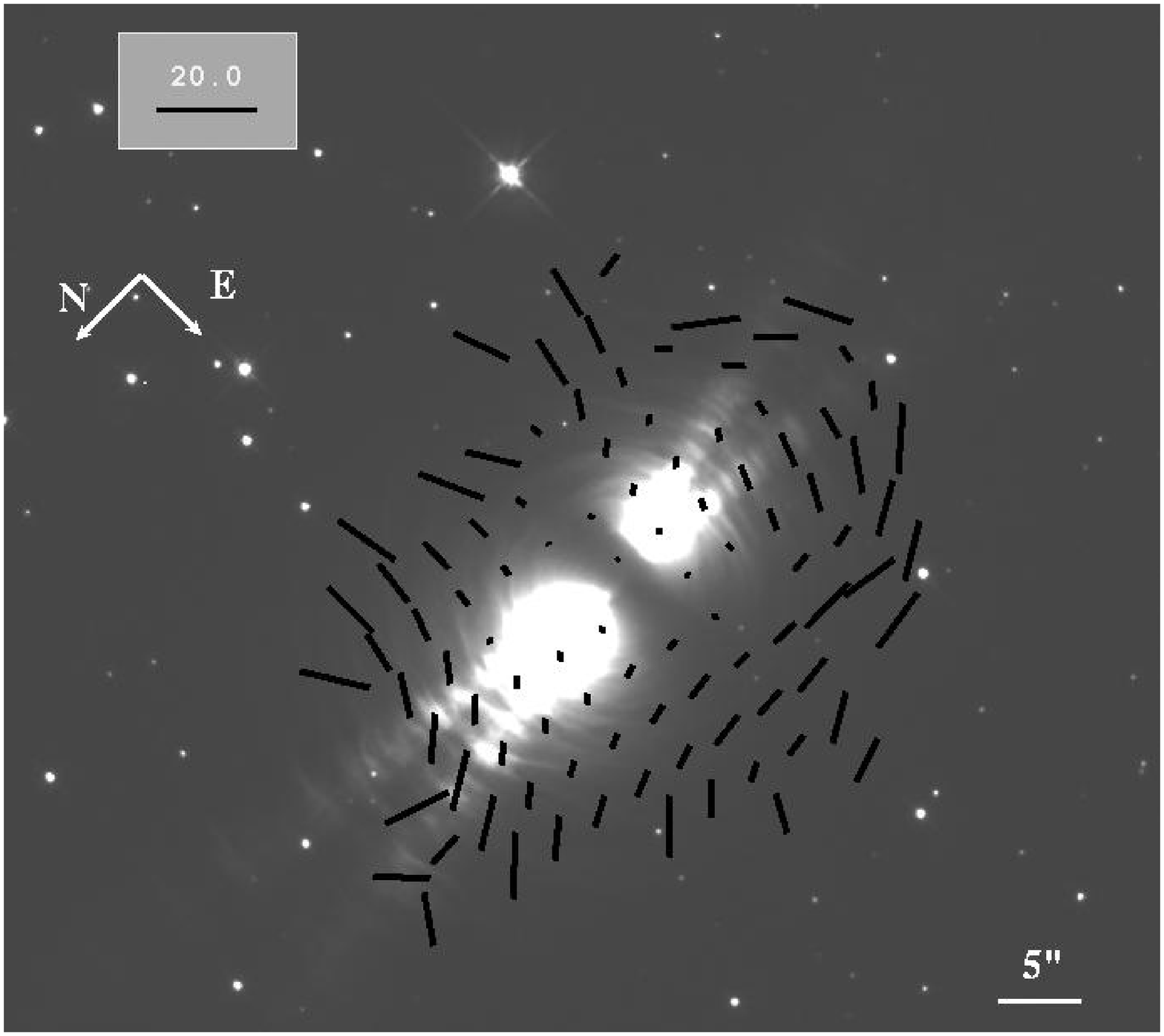}}
{\includegraphics [height=7.5cm]{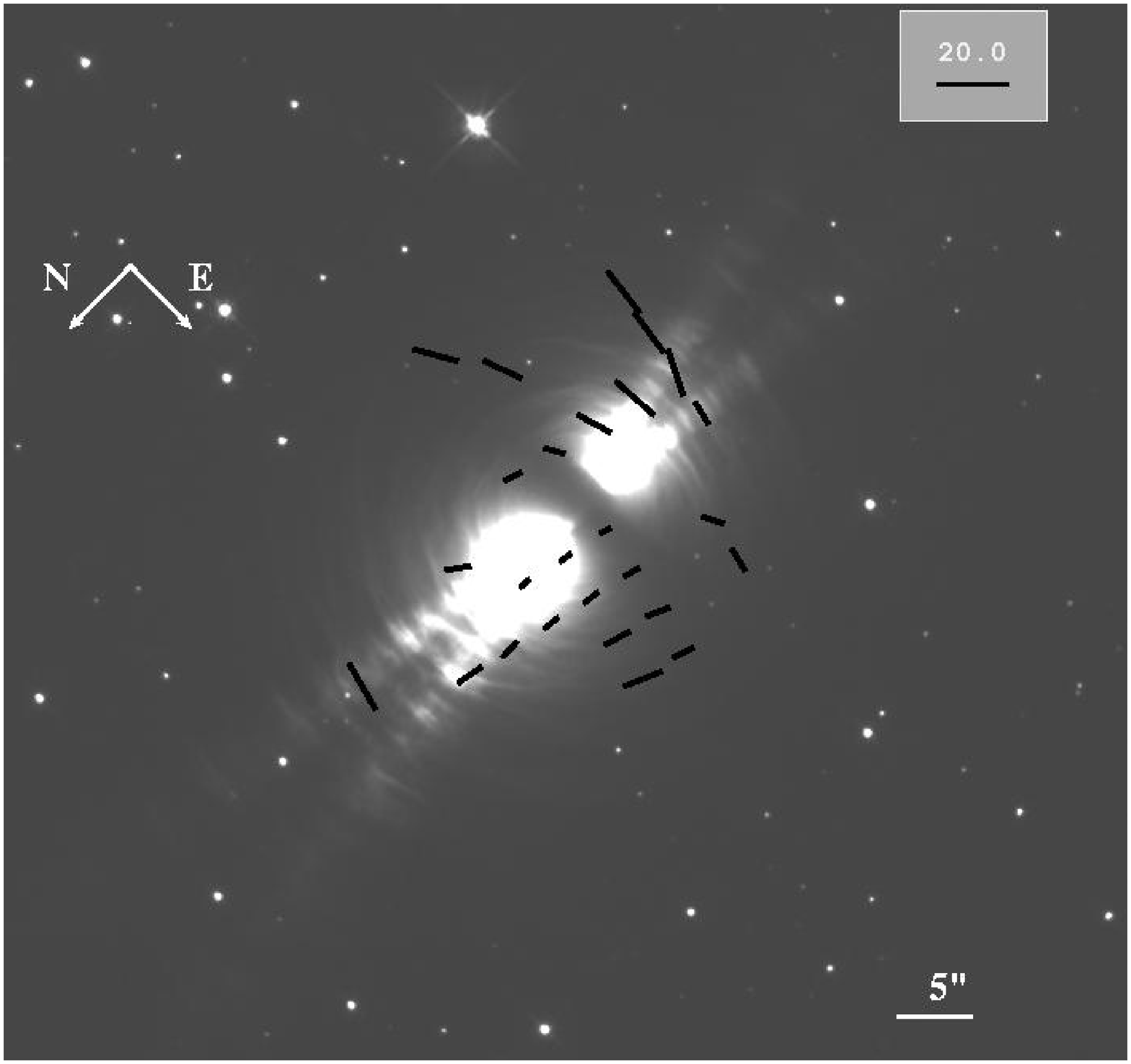}}
}
\caption{\label{}Scuba 850 and 450$\mu$m results on CRL2688.  North on 
the top and East on the left. The axes give the image scale in
arcseconds. {\it Top left panel}: The 850$\mu$m map of CRL 2688, with contours
from 1 to 5\% and 10\% of the peak. {\it Top right panel}: The 850-$\mu$m
vector polarization showing the magnetic field orientation. The general
outflow direction is indicated by A and the equatorial plane by B. The
polarization vector scale (showing the degree of polarization) is set at
10\%. {\it Bottom panels}: HST images (WFPC2, filter F606W) with the overlaid
magnetic field corresponding to the 850$\mu$m map (left) and 450$\mu$m map
(right). The dust continuum is elongated along the outflows. The field shows a
complicated structure with components along the outflow and along the
equatorial plane. The polarization vector scale (showing the degree of
polarization) is set at 20\% for both maps.}
\end{center}
\end{figure*}

The proto-planetary nebula CRL 2688 (or Egg Nebula) is a bright carbon-rich
bipolar object characterized by two pairs of searchlight beams superposed on a
reflection nebula. The origin of the light beams has been suggested to be a
very close stellar companion \citep{Sahai1998, Kastner2004}, or the presence
of dust layers reflecting the light of the central star \citep{Goto2002}. The
bipolar reflection nebula shows a dark equatorial lane where a large amount of
dust may by present.

The 850-$\mu$m is presented in Fig.~3-Top panels. It shows
elongation in the direction of the polar outflows. The equatorial extent is
less but at the lowest contour the torus is more extended on the eastern
side.  The extent is $\sim$60''$\times$40'' although it is not clear precisely
where the emission ends.  The bright core at 850$\mu$m shows a FWHM of
$16\times14$ arcsec$^2$, at a position angle of 25 degrees which is in the same
direction as the outflow. We could not measure the core elongation at
450$\mu$m due to the under-sampling.  The spatial resolution is insufficient
to separate the two light beams, but the width of the emission suggests that
the dust traces the larger lobes which the light beams illuminate, and that
these beams themselves are not present in the far-infrared data.  (The
elongation of the core is in fact along the line of one of the two beams only.)

CRL 2688 presents the highest number of polarization vectors in our
sample. For this nebula, the under-sampled 450$\mu$m map gives some
additional and useful indications (even if we observe a lack of vectors due to
this under-sampling).  The polarization vectors cover the full nebular extent
(seen at both wavelengths), as they do in NGC 7027. The degree of polarization
is not uniform and strongly decreases near the center. This phenomenon is more
visible in the 850$\mu$m map, which suggests that beam depolarization may play
a role.  At 850\,$\mu$m, the mean value of polarization of the region
containing the bipolar lobes is about 3.2$\%$, the outer region has a mean
value of 8.8$\%$, and the region of the dark lane shows a mean degree of
polarization of 1.4$\%$.

If we draw a line passing through the nebula in a longitudinal direction
(NNE-SSW fig.3 bottom-left panel), we can see distinct behaviours of the
magnetic field on either side of this line: on the eastern side, the
orientation of $\vec{B}$ is mainly in the direction of this line, while on the
western side, the magnetic field appears perpendicular to it.

The 450-$\mu$m map (Fig.~3-bottom-right panel) shows the magnetic field at
higher resolution, confirming the bimodal distribution. This map is
under-sampled and only some positions in the nebula are covered. The map
gives a suggestion of a superposition of a toroidal field and one aligned with
the polar outflows. The field becomes less ordered towards the tip of the
outflow direction, but the emission here is faint and the uncertainties on the
polarization vectors are larger.

The complicated magnetic morphology makes it likely that the dynamics have
shaped significant parts of the field, rather than the magnetic field shaping
the nebula. The structure may show the superposition of two components over
most of the area of the source. The lack of polarization in the centre
suggests that the bright core is not polarized, or its polarization is
averaged out over the JCMT resolution.  The outermost field vectors in the
longitudinal direction are along the outflow, and this may indicate a field
carried along by the outflow. This is most apparent in the north when looking
at the outer vectors direction in the figure 3-bottom-left, in the south the
outer vectors point towards the corresponding outflow so we assume that the
magnetic field may be carried by this outflow.

The grain alignment, which is perpendicular to the direction of the magnetic
field, is in the direction of the outflows on the western
side. On the eastern side, the grain alignment is toroidal.

High resolution molecular images in the literature include IRAM CO J=2-1 data
\citep{Cox2000} and HST infrared H$_2$ images \citep{Sahai1998}.  The
continuum-subtracted H$_2$ image in Fig. 4, shows multiple jets, located both
in the equatorial plane and towards the polar directions. The CO data show
these jets to be the tip of flows originating much closer to the star. These
jets all fall within the Scuba 850-$\mu$m core, indicating that indeed
structure is present on scales much smaller than the JCMT beam size. Thus,
beam dilution and beam-depolarization of the magnetic field is likely.

We note that the elongation visible in the Scuba image, on the eastern side,
is in the same direction as the largest molecular subjet E2 in Fig.4, or A1 in
\citet{Cox2000} (but the Scuba structure is much larger). The polarization
vectors seem unaffected. On the western side, both the molecular and the Scuba
images indicate a smaller extent of the envelope.  The polar outflows as seen
in the molecular lines do not show the light beams, but instead show
relatively well-collimated lobes. The lobes are brighter in the north. The
polar lobes in the Scuba image are consistent with this, both in direction and
in brightness, but again are much larger than seen in the molecules.  The
molecular emission shows the wind-blown cavities, while the Scuba emission
arises in the surrounding shells.

\subsection{NGC 6302}

\begin{figure}
\begin{center}

{\includegraphics [height=7cm]{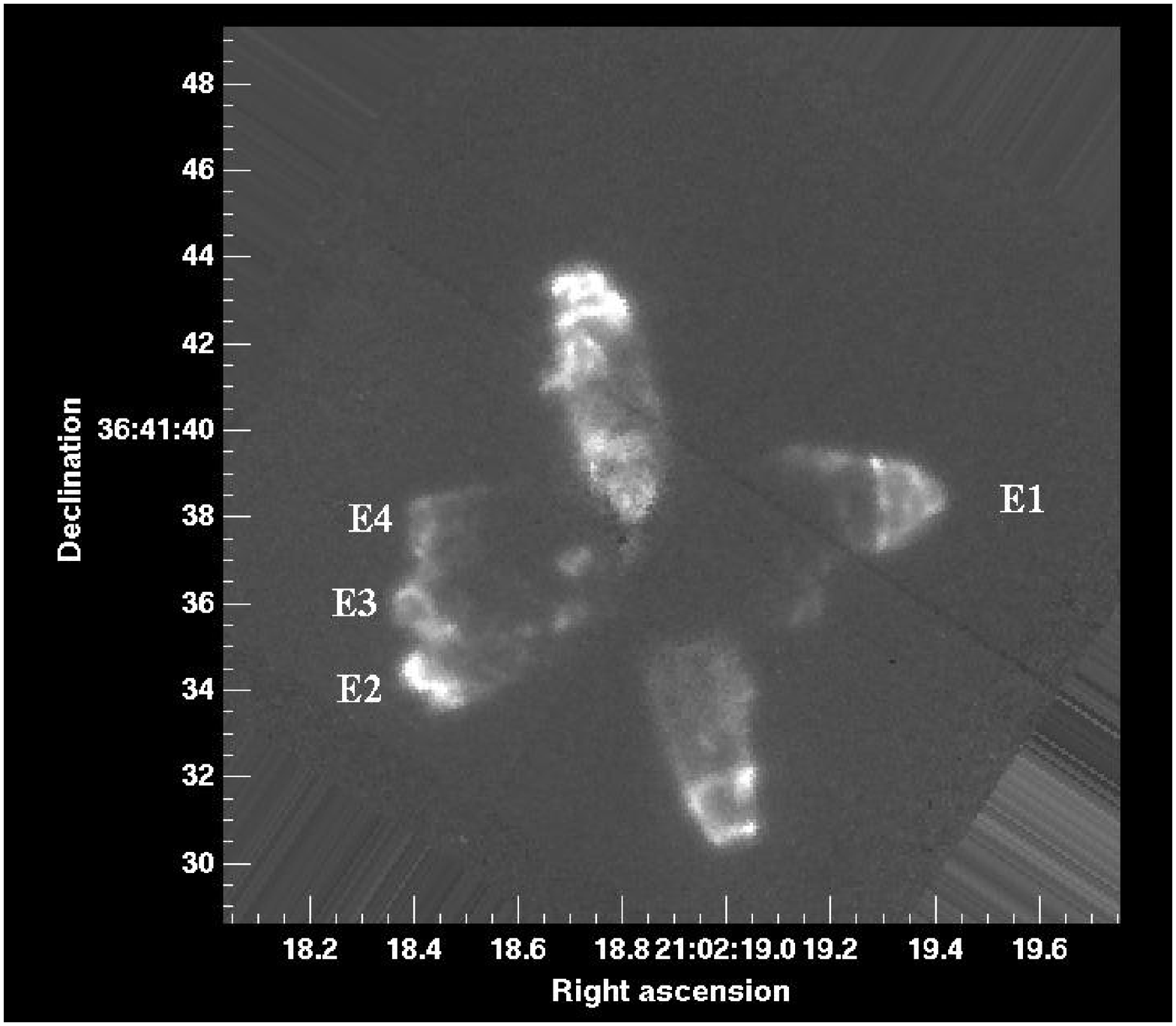}}

\caption{\label{} Continuum-subtracted H$_2$ (HST) image showing the jets in 
CRL 2688. [North at  the top and East at the left]. They are named as in 
Sahai et al (1998). The jets are all located within the compact 850-$\mu$m
source, and the some perturbations seen in the Scuba data can be related to the ``sub-jets'' E2 or A1.}
\end{center}
\end{figure}

\begin{figure}
\begin{center}
{\includegraphics [height=7cm]{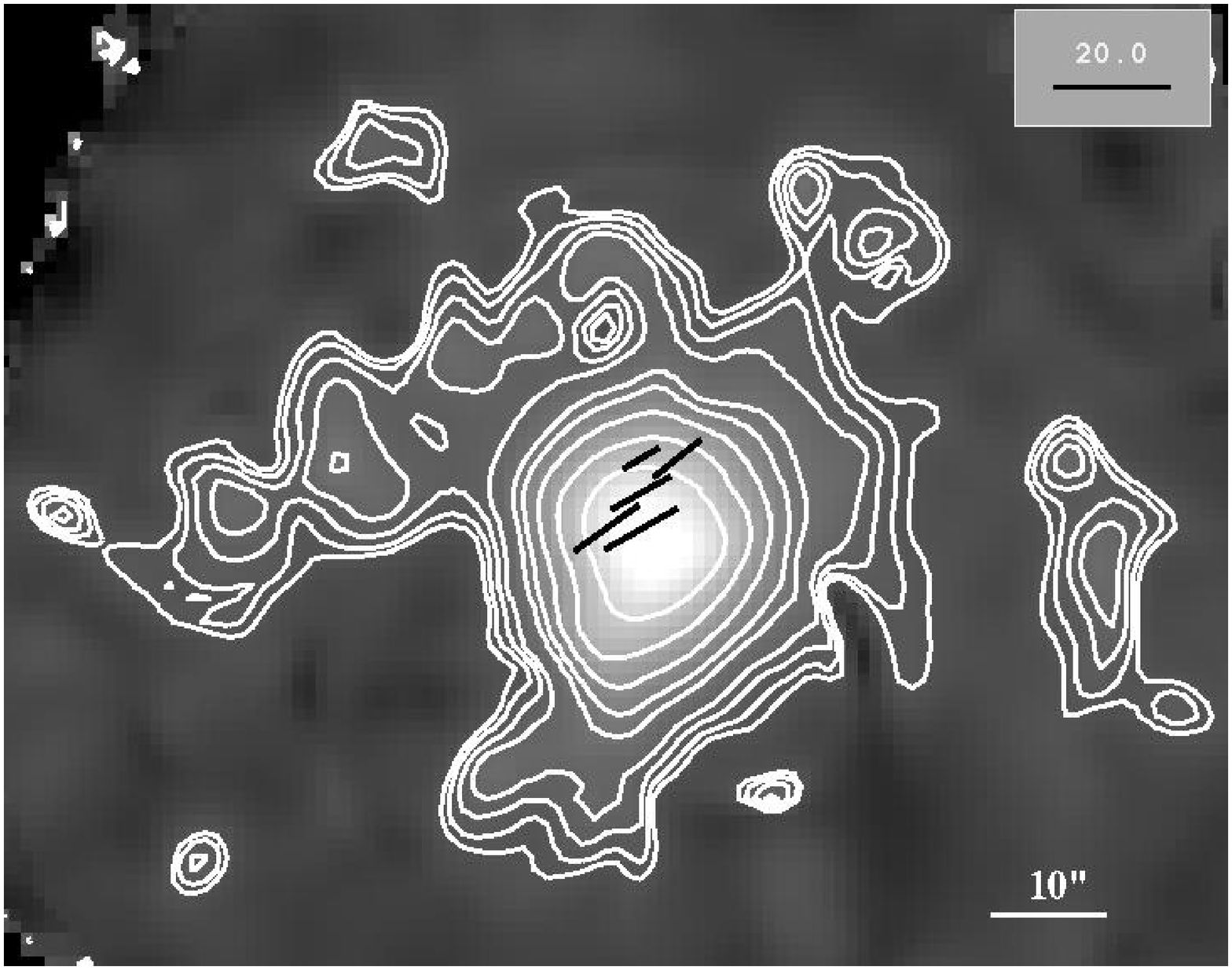}}
{\includegraphics [height=6cm]{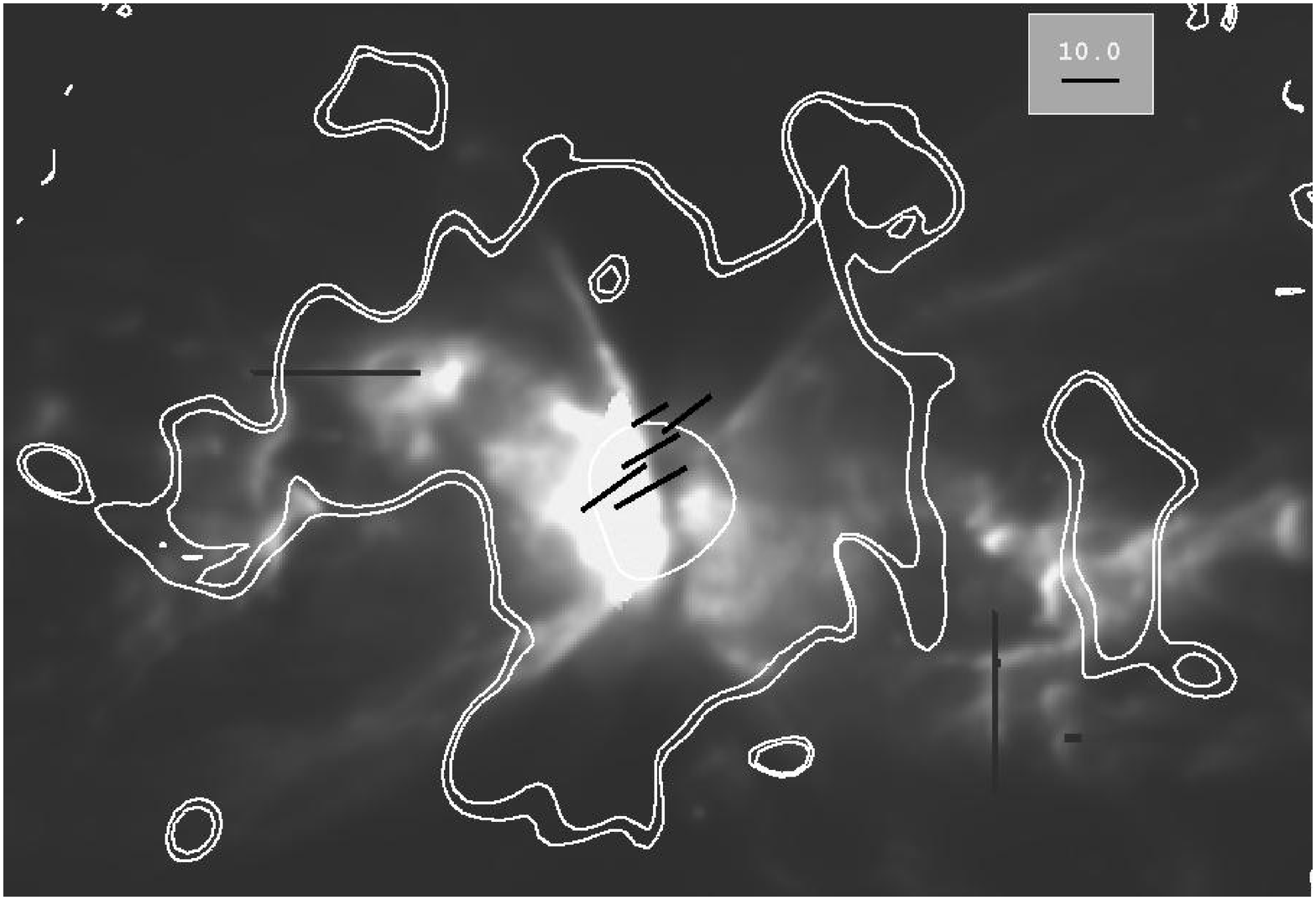}}
\caption{\label{}Scuba 450$\mu$m results on NGC 6302. North on 
the top and East on the left. {\it Top panel}: The map shows the magnetic
field orientation on the planetary nebula with contours from 1 to 5\%,
10\%,20\% and 25\% of the peak. The polarization vector scale (showing the
degree of polarization) is set at 20\%. Note that the arcs seen 20 arcsec out
are part of the beam side lobes of the very bright core. {\it Bottom panel}:
450$\mu$m Scuba map of NGC 6302 overlaid on the F656N band HST image
(WFPC2). North on the top and East on the left. The external hexagonal
contours do not belong to the star but are representative of the
bolometers. The main Eastern outflow seen on the submillimeter image coincides
with the one observed with the HST. Note that the arcs seen 25 arcsec out are
part of the beam side lobes and the two long lines (horizontal and vertical) on both sides of the nebula, are artifacts from the HST image and not polarization vectors.}

\end{center}
\end{figure}

This oxygen-rich planetary nebula \citep{Pottasch99} is a proto-typical
butterfly-type nebula.  It shows two extended lobes that result from a bipolar
outflow, with a dark lane in the center. It has been studied by
\citet{Meaburn1980}, \cite{Meaburn2005}, \citet{Matsuura2005b} 
and \cite{Casassus2000}.

The Scuba observations were optimized for 450-$\mu$m. They show a bright core,
of FWHM $14\times12$ arcsec$^2$ (larger than the beam), with the long axis in
the north-south direction, with an extension towards the south-southeast
(Fig.5). We underline the fact that the long horizontal line in the eastern part of the nebula and the long vertical line in the South-West (Fig.5 bottom panel), are artifacts from the HST image and not Scuba polarization vectors. The arcs seen at 25 arcsec are likely part of the diffraction pattern
from the very bright core.  The polarimetric data obtained at 450$\mu$m shows
only five polarization vectors. They do not line up with either
the dark lane or the outflow. However, they are  fairly well aligned with
the ellipsoidal radio source in the center of the nebula. The radio nebula
shows the inner ionized region, confined by the dense torus: the elongation is
perpendicular to the torus. 

The difference in position angle of the inner torus with the dark lane is
interpreted as a warped disk \citep{Matsuura2005b}.  Further from the center
the outflows have different position angles, and eventually become east-west
in the outer bipolar lobes. The 450-$\mu$m polarization indicates that the
magnetic field may be oriented in the direction of the inner outflow.

We have no data for magnetic fields elsewhere in the  outflows 
even at 850$\mu$m. 

\begin{table}
\begin{tabular}{|c|c|c|c|}
\hline 
Name&  Band ($\mu$m) & Deg ($\%$) & Angle (deg) \\
 \hline
 NGC 6537 & 850 & 11.2$\pm$2.2 &  26.5$\pm$5.7 \\
NGC 6302  & 450 & 11.4$\pm$1.6 &  32.7$\pm$4.6 \\
NGC 7027  & 450 &  8.2$\pm$0.9 & -18.5$\pm$3.7 \\
CRL2688   & 850 &  6.8$\pm$1.0 & -22.6$\pm$4.5\\
\hline
\end{tabular}
\caption{\label{} Mean values of the polarimetric parameters for the 
Post-AGB objects,  Deg: degree of polarization, Angle: position angle of the 
polarization. These values concern the main band studied for each nebula.}
\end{table}

\section{Discussion and Conclusion}

The sub-millimeter observations of the four post-AGB objects reveal new
information regarding the distribution of the magnetic fields, the dust
emission and the link between the two components. 

Extended dust emission was seen for all four nebulae. For CRL 2688,
the large polar lobes have been detected. The orientation of its polar
lobes is along  only one of the two light beams. The equatorial emission is
also extended, and shows an asymmetry which is correlated with the appearance
of the (much more compact) equatorial jets. These jets originate close to the
star and it is unlikely that they are affected by structures at much larger
scales. Instead, the jets may affect the dust emission through heating.
In NGC 6537, an asymmetry in the dust correlates with differences in the
extinction map, suggesting that here the Scuba map shows a true asymmetry
in the dust distribution.

The presence of detectable polarization shows that at least some of the 
dust grains are not spherical.

\subsection{Targets}
Polarization is detected for all four nebulae. This in itself suggests that
magnetic fields are common for these types of objects. It is thus important
to discuss what types of objects were selected.

First, the four targets are all bipolar nebulae. The least pronounced
morphology is shown by NGC 7027, but even here the molecular maps clearly show
the underlying bipolar structure. We do not have information on the existence
of magnetic fields in less bipolar (e.g. elliptical or spherical) nebulae.

Second, although three of the nebulae have very hot central stars and ionized
cores, all have extended molecular envelopes. It is argued above that the
detected fields are located in the molecular or atomic regions. The presence
of molecular envelopes around ionizing stars indicates very dense nebulae, and
therefore very high mass loss rates on the AGB.

Third, the dense nebulae with relatively small ionized regions imply that
their stars have evolved to high temperatures before the nebulae have had time
to expand significantly. Such fast evolution is characteristic for high-mass
central stars. The cooler object (CRL 2688) may have a lower mass, but the
fact that it has become a carbon star suggests it still has a relatively
higher initial mass ($M_i \gsim 3 M_\odot$), as lower-mass stars experience
insufficient dredge-upo to become carbon rich. For NGC\,6302, a progenitor
mass of 4--5\,M$_\odot$ has been suggested, based on the mass of the
circumstellar envelope \citep{Matsuura2005b}. The precise initial masses are
not well determined, but high mass progenitors appear likely for all four
objects.

Thus, the result of the survey can be interpreted that magnetic fields tend to
be present for high mass progenitors evolving into bipolar nebulae, and that
the fields are detectable while the nebulae still have molecular envelopes.

\subsection{Field location and orientation}

Overall, we see fields aligned with the polar directions and/or toroidal
fields in the equatorial plane. Assuming that the dust in the polar
flows originates from the dust reservoir in the disk or torus, this
suggests that the field also originates there. The alignment with the
polar flow is interpreted that the gas carries the field with it.

NGC\,6302 is the only source in our sample without evidence for a toroidal
field. This nebula has a complicated morphology with a warped disk and the
nebula is oriented at 45 degrees with respect to the equatorial dust lane. It
is therefore difficult to ascertain the alignment of the field with any
particular morphological component. However, alignment with the radio core
seems most likely. 

The other nebulae show evidence for toroidal field components. We therefore
suggest that the initial configuration of the field is toroidal, and is located
in the equatorial torus. 

The polarization in all cases appears to be seen in the neutral/molecular
regions. Towards the central, ionized regions, the percentage of linear
polarization is much reduced. This may in part be because of the beam
averaging over regions of different polarizations, but the ionized regions may
also be expected to lose a directional field (non-constancy of $\vec{B}$). The
reason for the absence of a detectable magnetic field in the ionized region
can also be explained by the lack of dust (so $\vec{B}$ is not carried). For
NGC\,6302, the field is aligned with the ionized core but is likely located
around rather than within the core.

Polarization will  not be detected if the field is oriented along the line
of sight. This is expected for a toroidal fields near the tangential points of
the torus.  This may be present especially in CRL 2688, where two components
are found at different locations.

\begin{figure*}
\begin{center}
{\includegraphics [height=5.5cm]{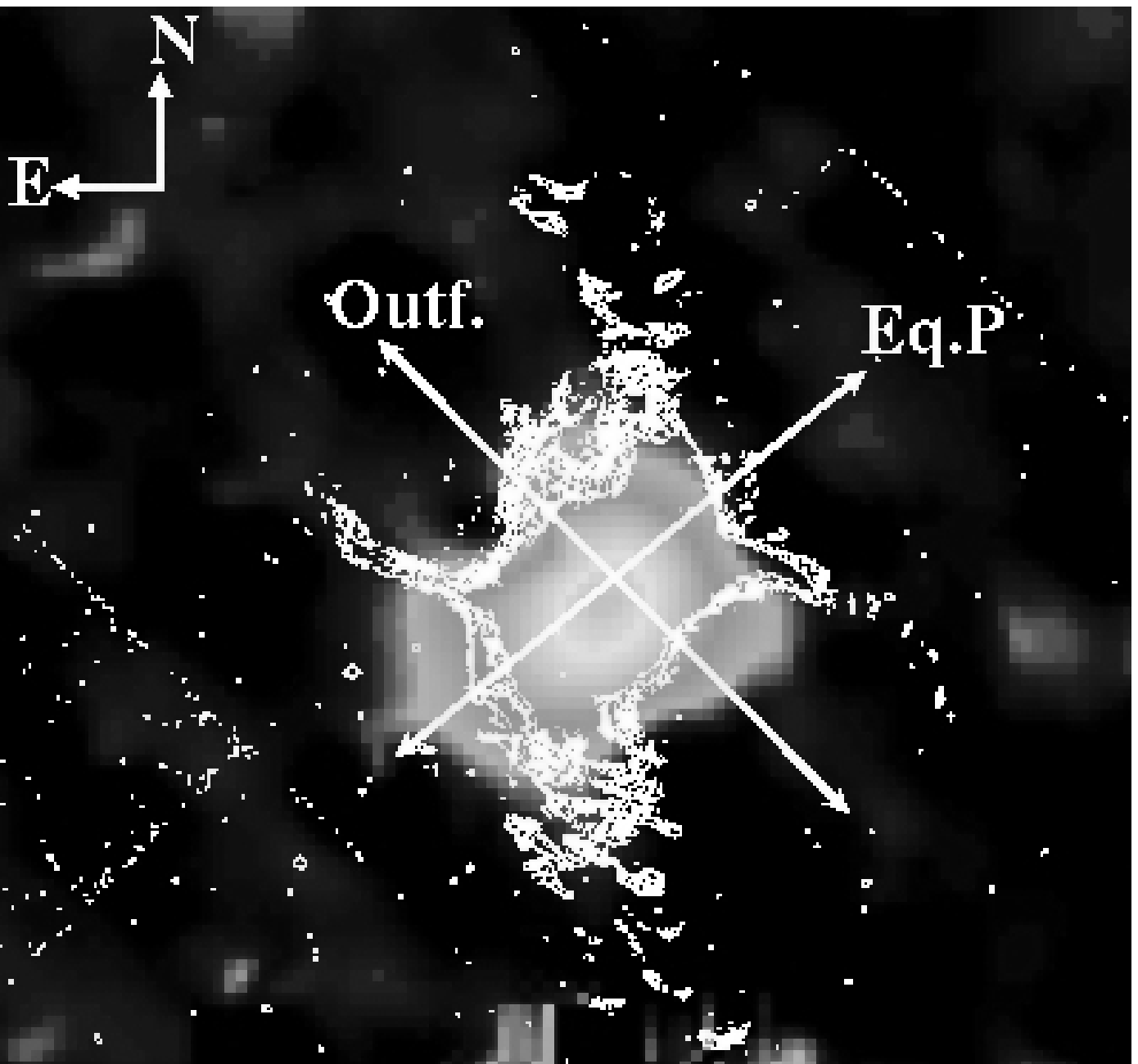}}
{\includegraphics [height=5.5cm]{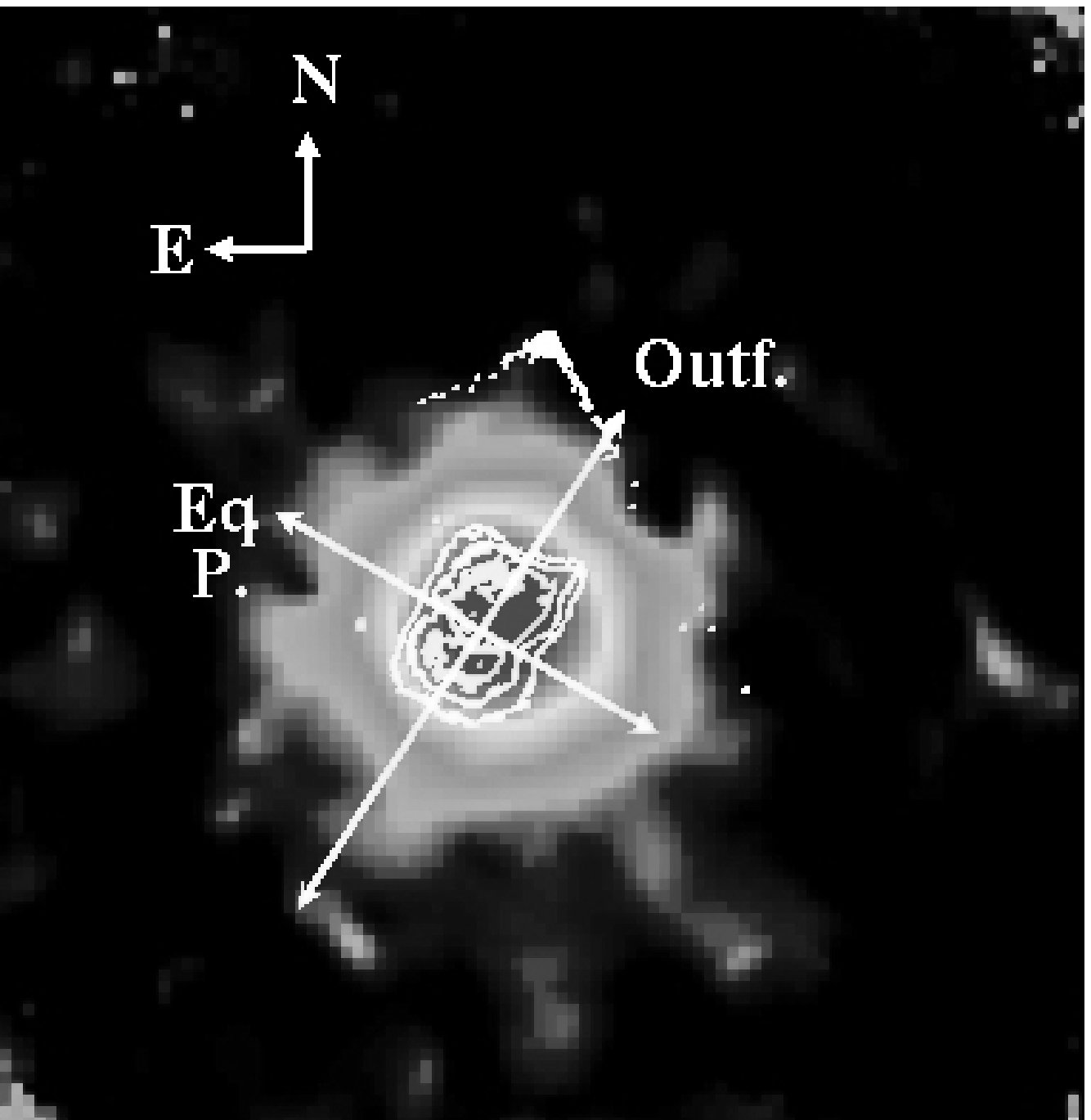}}
{\includegraphics [height=5.5cm]{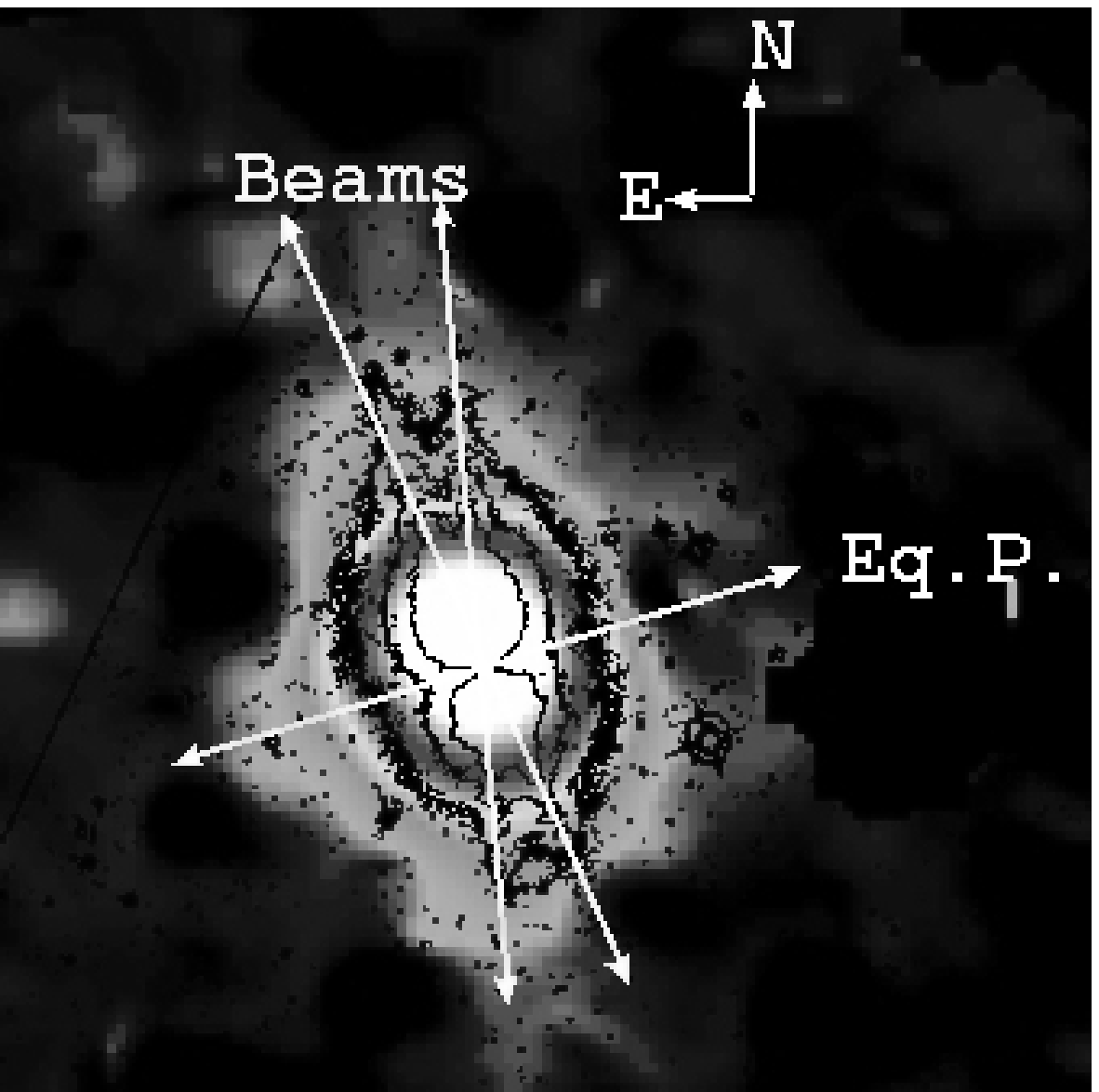}}
\caption{\label{}. Location of the Scuba dust emission, compared to the
 HST images (contours). We present respectively NGC 6537, NGC 7027 and CRL 2688. NGC 6302 is shown in Fig.5 (bottom panel).}
\end{center}
\end{figure*}
\subsection{Chemistry}

The sample includes two oxygen-rich nebulae (NGC 6537 and NGC 6302, both
PNe with hot central stars) and two carbon-rich nebulae (NGC 7027 and CRL
2688: the latter is the cooler post-AGB nebula in our sample). This sample
is too small to draw general conclusions. However, we point out the following
points of interest, for future study.

\begin{itemize}
\item
The O-rich PNe both present collimated magnetic fields concentrated near the
central star. In both cases, we have an organized magnetic field in the
central region, resistant to distortion effects, but do not detect more
distant fields.
\item
The two carbon-rich nebulae show extended fields covering their entire nebula.
In both nebulae, the magnetic field is more disorganized
than in the two O-rich stars.

\end{itemize}

The question whether there is a chemistry-related origin to these behaviours
is valid, but cannot be answered with the current sample. Dust chemistry could
play a role.  First, carbon grains tend to be smaller than silicate
grains. Larger grains may be more spherical, which would give less (or no)
polarization.  Both oxygen-rich nebulae show evidence for crystalline ice
\citep{Molster2002} (strongest in NGC 6302), indicating coatings on the
grains.  Second, amorphous carbon grains may be intrinsically
non-spherical, build up from graphitic carbon sheets.

For the O-rich nebulae, it is possible that only the hotter dust, above the
ice evaporation temperature, leads to non-spherical grains and a polarized
signal.  Thus, a difference in polarization does not necessary imply a
difference in magnetic field structure.

Both O-rich nebulae show evidence for weak PAH emission in their central
regions. The origins of the PAHs is not clear, but they are seen only in
irradiated regions. The continuum radiation from the dust component
containing the PAHs is only seen at short wavelengths (warm dust) and
contributes very little to the sub-mm flux \citep{kemper2002, Matsuura2005b}.
Thus, the polarization signal is seen in the oxygen-rich grains, not the PAHs.

Iron needles could be considered as carriers of the polarization: in O-rich
stars, some or most of the iron is incorporated in the silicate grains
\citep[olivines and pyroxenes:][]{Ferrarotti2001}, but for a C/O ratio close 
to and above unity, solid iron and FeSi become more important
\citep[e.g.,][]{Ferrarotti2002}. Metallic, non-spherical iron grains may
contribute to the NIR opacity in high mass-loss OH/IR stars
\citep{Kemper2002b}.

\subsection{Comparison  less evolved/ more evolved nebulae}
 
The extent of the nebulae shows that the objects differ in age. The nebulae of
CRL 2688 (Post-AGB) and NGC 7027 (young PN) are younger than NGC 6537 (PN) and
NGC 6302 (PN). (The lobes of the last object extend to over 2 parsec across.)

 Our submillimeter polarimetric data show that the magnetic field is better
organized for the older nebulae. Assuming an originally toroidal field, later
carried by the bipolar flows, the best organized field is expected if one
component dominates. The bipolar flows are most massive early in the post-AGB
evolution \citep{Bujarrabal2001}. As they diminish, the remaining field may
become less confused.  However, evolving dust characteristics may also play a
role: dust grains become larger in older disks, and may show less polarization
as a result.  It is not possible with the current sample to separate
evolutionary and chemical effects, but both are expected to occur.

\subsection{Field strength and origin}

The detected polarization contains information on the direction of the field,
but not its strength. A method to obtain the field strength out of
fluctuations in polarization was suggested by \citet{Chandra1953}.  Applying
this method to our data leads to field strengths of the order of mG.  This
equation assumes that the magnetic field dispersion is caused by Alfven
waves. The applicability to our sources may be in doubt, as the dispersion may
be dominated streaming motions.

In neutral-dominated media, Alfven waves contribute if the collision times
between ions and neutrals are shorter than the period of the Alfven wave. This
is the case for waves longer than

\begin{equation}
 \lambda_A > \rm  \left( \frac{0.3}{pc}\right) \left(\frac{B}{0.1\,mG} \right)
   \left( \frac{10^4}{n}  \right)^{3/2} \left( \frac{10^{-7}}{f_i} \right)
\end{equation}

\noindent
where $n$ is the particle density per cm$^3$, and $f_i$ is the fractional
ionization \citep{Hildebrand1996}.  For our sources, $f_i$ may be high except
in the densest regions of the tori. Assuming a typical source size of
0.05\,pc, and a field strength of a mG, Alfven waves may contribute if
$f_i>10^{-4} $. This is a plausible value for the outflows. 

Previous observations of OH Zeeman splitting has shown field strengths in
OH/IR stars and post-AGB stars of a few mG at $3 \times 10^{15}$ -- $2 \times
10^{16}\,$cm from the stars \citep{Bains2004, Bains2003}. Much larger values
have been inferred from water masers: several Gauss at $\sim 2 \times
10^{14}\,$cm \citep{Vlemmings2005}. The difference is consistent with a dipole
field, which gives an $r^{-3}$ dependence, and less consistent with a
solar-type field $r^{-2}$. Water masers trace high density clumps and the
measured field strengths may not be fully representative of the surrounding
areas.

 Assuming we look at typical distances of 5 arcsec at 1 kpc, our measurements
are at $\sim 5 \times 10^{16}\,$cm from the stars. Compared to the OH and
water masers, one would expect a field of 1mG or less at this distance.  This
suggest that the direction changes in the field are due to streaming motions,
rather than Alfven waves.

The grains become aligned with the magnetic field with the Davis-Greenstein
mechanism. The time scale for this to occur is \citep[][Ch. 11]{Krugel2003}
$$ t_{\rm rel} \propto B^{-2} $$ and, for fields of the order of a few mG, is
typically $10^6$\,yr. The age of the nebulae is $\sim 10^4$\,yr. For this time
scale, the alignment of the dust with the magnetic field requires fields in
excess of those seen from OH maser emission. The alignment is therefore likely
to originate close to the central star, at $r<10^{15}\,\rm cm$, and is
maintained while the nebula expands.

\subsection{Evolution}

For OH/IR stars, evidence for dipole fields has been presented by, e.g.,
\citet{Vlemmings2005}. For more evolved stars, in the early post-AGB evolution,
the presence of both toroidal (IRAS 20406+2953) and poloidal (OH17.7-2.0)
fields are inferred from OH observations \citep{Bains2004, Bains2003}.  The
poloidal fields arise from a stretched dipole. For the even later evolutionary
stage studied here, a strong toroidal field component is found, combined with
a poloidal field.

A toroidal field can form out of a dipole field by rotation. This already
makes it likely that the formation of the torus and the formation of the
toroidal field are related. A binary companion can be the source of the
required angular momentum. Rotation of the star itself is less likely,
as its angular momentum can be expected to be lost early on in the mass loss.
A binary companion, on the other hand, can deposit its angular momentum
during the time of the peak superwind, at the end of the AGB mass loss.

The fields detected via OH and H$_2$O masers have been claimed to be
strong enough to dominate the dynamics of the nebula
\citep{Bains2004,Vlemmings2005}. \citet{Soker2006} argues that fields of
this magnitude cannot be generated by the star itself, and should be
attributed to a companion. Based on this, the basic shaping mechanism of
the nebula is found in binarity. However, once the strong fields have been
generated, they would be a significant factor in the further evolution of
the nebula towards the PN phase.

\subsection{Summary}

We present the discovery of magnetic fields in four bipolar post-AGB stars
(NGC 6537, NGC 7027, NGC 6302 and CRL 2688).  This confirms the earlier work
of \citet{Greaves2002} for two of these.  The fields are mapped at high
resolution (for sub-mm), using either 450 or 850$\mu$m.  The sub-mm emission
traces the extended emission. In CRL 2688, we find evidence for the polar
lobes which contain the well-known search beams.

All objects show polarization indicative of grain alignment by magnetic
fields. Toroidal fields are found for three objects, and poloidal fields for
two (CRL 2688 shows evidence for both). The alignment of the field with nebula
is least certain for NGC 6302, where the nebula shows a multipolar structure.
Our results suggests that magnetic fields are common in these types of
targets: bipolar nebulae with intermediate-mass progenitors. The fields are 
long lived as they are observed over different evolutionary stages. 

The data also show evidence for elongated grains. The polarization is
stronger and more extended for the carbon-rich nebulae. This may show that
amorphous carbon grains are intrinsically non-spherical, e.g. the sheet-like
structure of graphite. The oxygen-rich nebulae show polarization only in their
central regions. This is interpreted in that the dust grains in the torii are
larger and more spherical.

The poloidal fields in the polar flows suggest that here the field is carried
along by the flow and that these flows are not magnetically confined. The
toroidal components in the torus may be more important, for the dynamics. As
the fields in AGB stars are dipole-like, the toroidal field structure appears 
to be a later phase of evolution. We suggest that this transition occurs when
the equatorial field is wound up through the interaction with a companion.
Following \citet{Soker2006}, we suggest that the original shaping agent
is a binary companion. Once the field has been wound up, it may however 
become dynamically important for the subsequent nebular evolution.

\section*{Acknowledgements}

L.S. thanks Mikako Matsuura for many useful comments and for providing the
reduced HST data.  The James Clerk Maxwell Telescope is operated by The Joint
Astronomy Centre on behalf of the Particle Physics and Astronomy Research
Council of the United Kingdom, the Netherlands Organisation for Scientific
Research, and the National Research Council of Canada.

\bibliographystyle{mn2e}

\bibliography{paper_scuba_az_rev}

\end{document}